# Frictional Stabilities on Induced Earthquake Fault Planes at Fox Creek, Alberta: A Pore Fluid Pressure Dilemma

L. W. Shen[1,3]†, D. R. Schmitt[2], and R. Schultz[3]

[1] Institute of Geophysical Research, Department of Physics, University of Alberta, AB, Canada

[2] Department of Earth, Atmospheric and Planetary Science, Purdue University, West Lafayette IN, USA; formerly at Institute for Geophysical Research, Department of Physics, University of Alberta, AB, Canada

[3] Alberta Geological Survey, Alberta Energy Regulator, AB, Canada.

†Corresponding author: Luyi W. Shen (luyi@ualberta.ca)

**Key Points:**

- The slip-tendency along the fault planes are determined using a quantitative model for the stress tensor at the hypocentral locations.
- The faults cannot be stable at the ambient high pore fluid pressures within the Duvernay Formation.
- Triggered planes of weakness need not be perfectly optimally oriented to the stress field.

**Abstract**

Earthquakes induced during hydraulic fracturing operations have occurred in a number of locales. However, in-situ studies aimed to discern the triggering mechanism remains exclusively statistical in their nature. Here, we calculate the fault slip-tendencies of eleven hydraulic fracturing induced earthquakes in a historically aseismic area using a recently constructed quantitative model for in-situ stresses. It is shown that the ambient pore pressures of the nearby Duvernay unconventional reservoirs can provide enough $\Delta P_f$ triggering fault movement. The local fluid pressures acting on the fault could readily be increased above the critical value if a hydraulic connection exists between the fault and a propagating hydraulic fracture. The critical pressures necessary to induce slip, is estimated using a probabilistic model that incorporates uncertainties of stress and fault mechanical properties. These critical pressures are greater than expected hydrostatic pressure but less the pore pressures of nearby unconventional reservoirs.

**Plain Language Summary**

Hydraulic fracturing operations within the Duvernay Formation, a major hydrocarbon source rock for the Western Canada Sedimentary Basin, have been linked to induced earthquakes primarily on the basis of statistical analysis. Here, we attempt to examine whether faults in the area are prone to slip, casing an earthquake, using well constrained values for the state of stress in the earth at the locations of the induced earthquakes. Perhaps most interestingly, the fault planes along which slip occurred could not have remained immobile if the fluid pressures directly acting on the faults is the same as the anomalously high pore fluid pressures measured within the Duvernay Formation itself. However, the faults have been historically quiescent and this suggests that the fluid pressures acting on them must be lower. This might be accomplished if pressures are relieved by fluid migration to overlying hydrocarbon reservoirs. Further, this suggests that the earthquakes are triggered by elevated fluid pressures communicated to the pre-existing faults.

1. Introduction

The use of hydraulic fracturing (HF), particularly from multiple stages along horizontal boreholes, to access hydrocarbons from low-permeability unconventional and geothermal reservoirs continues to accelerate; and as such it becomes increasingly important that the risks associated with such practices are understood. Essentially, a HF operation consists of rapidly pressurizing a section of a borehole create and propagate new fractures and to disturb natural fracture networks to increase the effective permeability, thus allowing hydrocarbons to flow back to the borehole to be produced. Numerous microseismic events ($M_W < 0$) result, with their locations useful in tracking fracture network growth; but the process by itself in insufficiently energetic to produce larger felt events. Despite this, some larger rare events with $M_W$ possibly up to $M_W$ 4.7 have been temporally and spatially linked to HF operations in Western Canada, the United Kingdom, the United States, and China [*Li et al.*, 2019; *López-Comino et al.*, 2018]. Further study of these events assists in assessing risk levels and may also help understanding of the nucleation of natural earthquakes.

These HF-linked events differ from the significantly greater numbers of earthquakes associated with broader long-term fluid disposal [*Foulger et al.*, 2018]. The circumstances under which they initiate, however, are not universal with some occurring during HF-operations either during either pumping to extend the fracture network or 'flow-back' as the fracture fluids return to the borehole, or some-time after the operations have ceased. The actual hypocenters of these events with respect to the known injection points, too, remain uncertain making interpretation of the trigger mechanisms difficult. These relationships in time and the ambiguities in location have led to authors variously argue that the events are triggered on pre-existing planes of weakness by either increasing the Coulomb shear stress by poroelastic stress transference [e.g., *Deng et al.*, 2016] or by decreasing the effective stress through transmission of fluid pressures via diffusion or direct hydraulic connection [e.g., *Shapiro and Dinske*, 2009]. However, the true nature of the faults, the fluid communication pathways, and the poroelastic properties of the materials in situ remain unknown, and these hypotheses remain speculative until information only accessible through drilling is obtained.

We do not claim to fully answer this problem here, but we do provide new information on the stability of the fault planes for a number of well-studied events linked to HF operations in the Duvernay Formation in the vicinity of Fox Creek, Alberta. Our analysis relies on a recently developed quantitative model for this area that accurately predicts the state of stress, including pore pressures, at each hypocenter [*Shen et al.*, 2019] subsequently allowing the Mohr-Coulomb stability of each fault plane to be fully evaluated. An unexpected finding is that the fault planes cannot be stable if the pore fluid pressure acting on them is the same as that measured within the reservoir. Conversely, the fluid pressures acting on the faults must be substantially reduced from those within the reservoir rock if the faults are to remain stable, as they appear to have been at least through the historical record prior to 2015.

Below, we begin with a brief review of the Mohr-Coulomb stability criterion used and of the circumstances associated with the Fox Creek events. The stress model is then applied to evaluate the fault stability at the locations of each HF-linked event and this analysis then extended to delimit the pore fluid pressures required for the faults to become unstable. This analysis leads to suggestions as to the reasons why pore pressures may be diminished on the faults and possible implications for migration of hydrocarbons from the prolific Duvernay source rocks into the overlying siliclastic conventional reservoirs.

## 2. Slip Tendency Analysis

The dynamic behaviour of fault slip is expected to follow empirical rate-state friction laws [*Marone et al.*, 1990; *Ruina*, 1983]. However, the earthquake nucleation still remains poorly understood [*Gomberg*, 2018]. It is often assumed that slip initiates on a pre-existing plane of weakness once the magnitude of its resolved shear traction $|\tau|$ exceeds the frictional restraint, which is proportional to the product of the effective normal traction $\sigma$. Usually, simple Coulomb friction based criteria such as the widely employed

Coulomb failure stresses [*Harris*, 1998] or slip tendency *SNR* [*Morris et al., 1996*] are used to assess the risk that slip will occur. Slip initiation is assumed to be regulated by the static-friction co-efficient $\mu$, the cohesion $C$, and the pore fluid pressure $P_f$ on the slip plane through the Terzhaghi effective stress law for shear failure. Accordingly, slip initiates when the ratio of the fault shear $|\tau|$ to normal $\sigma$ tractions (the shear to normal ratio *SNR*) resolved onto the fault plane overcomes the fault friction $\mu$:

$$\mu < \frac{|\tau| - C}{\sigma - P_f} \equiv SNR \qquad . \qquad (1)$$

Usually, the cohesion $C$ of an already-existing fault plane is assumed small relative to the stresses and is often ignored in such analyses [*Scholz*, 2019; *Zoback*, 2010] including *Morris et al.* [1996] original study. despite evidence of fault healing that has been interpreted as a time dependent friction [*Dieterich*, 1972] or complications associated with separating cohesive and frictional effects [e.g., *Weiss et al.*, 2016] If the tectonic stress states that generate the resolved tractions $\tau$ and $\sigma$ as well as the $C$ remain unchanged, then slip along the fault may be triggered by increasing $P_f$, a concept that was first tested by injection of water to a producing oil reservoir at Rangeley, Colorado [*Raleigh et al.*, 1976].

## 3. Fox Creek, Alberta, Events

Over the Western Canada Sedimentary Basin, only a small fraction of HF stimulations has been linked to induced earthquakes with $M_W \geq 3$ [*Atkinson et al.*, 2016]. These HF induced earthquakes are geographically clustered, with no induced events detected from the nearby operations targeting the same geological units [*Schultz et al.*, 2017]. Clustering of the seismicity has been statistically related to high pore pressure $P_P$ gradients [*Eaton et al.*, 2018], local geological structure [*Schultz et al.*, 2016], or volumes of injected fluids [*Schultz et al.*, 2018]. Statistical correlations, however, cannot explain the proximate lack of seismicity nor resolve the physical mechanisms. The lack of knowledge of these processes limits the mitigation responses to 'traffic light protocols' during HF operations [*Shipman et al.*, 2018].

Here, we carry out stability analyses using high-quality FM solutions for 11 induced earthquakes (Fig. 1a, Table S1) linked to HF operations in the Duvernay Formation near Fox Creek, Alberta [*Schultz et al.*, 2017]. These analyses rely on a recently developed quantitative model for the full Andersonian tectonic principal stress magnitudes for the greatest $S_H$, least $S_h$ horizontal and the vertical $S_V$ compressions, as well the $S_H$ azimuth $\phi$ [*Shen et al.*, 2019]. Predictive values for each of these components (Fig. 1b-e) are obtained through geostatistical modelling using numerous borehole measurements within the Duvernay Formation (Fig. 1b). *Shen et al.* [2019] subsequently calculate a range of absolute $S_H$ [*Terakawa and Hauksson*, 2018] from the distribution of stress 'shape factors' determined from FM inversions [*Vavryčuk*, 2014]. The inversion algorithm further provides the three orthogonal principal stress directions that are consistent with both the observed strike-slip mechanisms and with the Andersonian hypothesis that one principal stress is vertical [*Shen et al.*, 2019]. When resolved onto the FM inferred fault planes (Table S1), these stresses allow us to calculate *SNR* distributions in order to carry out sensitivity tests on the factors affecting slip initiation with Eqn. 1. It is important to note that the pore pressure $P_P$ developed in this model is obtained from direct measurements in numerous boreholes over the area; as will become apparent later we distinguish this from the more general pore fluid pressure of Eqn. 1. Further, these measured $P_P$ are often more than double the normal hydrostat and are often more than 90% of the $S_h$ magnitude.

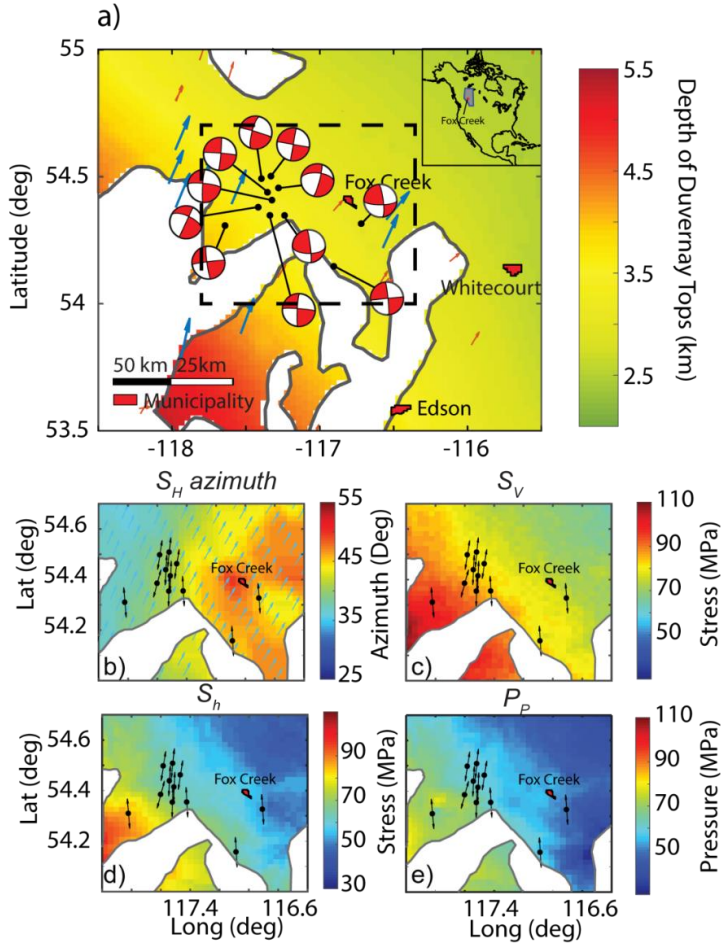

**Figure 1.** a) Focal mechanisms of the earthquakes analyzed in this paper. The black dots show the epicentres of each earthquake. The coloured background shows the depth of the Duvernay Formation that is cotemporaneous with Leduc reefs (white areas). Blue and orange arrows show the direction of $S_H$ determined by borehole observations from WSM [*Heidbach et al.*, 2016] and *Shen et al.* [2019]. Dashed line box indicates area of the directly measured stress tensor components shown as interpolated maps at the top of the Duvernay Formation in panels b-e. In these panels the black dots indicate the epicentres and the associated black arrows the slip direction determined from focal mechanisms. b) $S_H$ trend azimuth $\phi$ shown both as the colormap and on select grid points as cyan arrows c) vertical total compressive stress $S_V$, d least horizontal total compressive stress $S_h$ and e) formation pore pressure $P_P$. The greatest horizontal total compressive stresses are described by a statistical distribution at each grid point and are not shown (see supplementary information).

Two adjacent HF-induced earthquakes ($M_W$ 3.6, Jan 23, 2015; $M_W$ 4.1, Jan 12, 2016) are selected in part because of the availability of additional active-source seismic attribute images [*Chopra et al.*, 2017, Fig. S5] and careful determination of the epicentral locations and depths [*Wang et al.*, 2017]. There, FM indicates strike-slip motion on subvertical N-S striking fault planes. A red traffic light protocol was triggered for one of these events during HF operations [*Shipman et al.*, 2018] and the ground motion was locally felt. Various lines of evidence [*Eaton et al.*, 2018] suggest that the depth of the $M_W$ 4.1 event (and associated cluster) lies at ~3.5 km, coincident with HF of the Duvernay Formation [*Wang et al.*, 2017]. The lateral resolution of the stress model is ~2 km and the values for stress and $P_P$ (Fig. 1, Table S2) are nearly the same for both events under the assumption they occur within or close to the Duvernay Formation. The slip on these two events is well oriented with respect to the stress field (Fig. 2a-c) For the sake of comparison, the stability on a third fault plane associated with the Mw 3.9 (Jun 13, 2015) event that less optimally aligned with the stress field (Table S2) is also calculated (Fig. 2d-f).

Some description of Fig. 2 is necessary. Following *Morris et al.* [1996] a value for *SNR* is calculated on the entire set of planes possible with its value represented by a color positioned at the point that each plane's pole intersects the hemispherical stereonet projection. These plots are useful in evaluating the range of stable and unstable fault orientations. Further, the *SNR* in Fig. 2 is calculated (Eqn. 1) assuming $C = 0$, using the most probable $S_H$, and with three options of $P_f$ in which it is either omitted, made equal to the normal hydrostatic pressure expected, or assigned the measured value for $P_P$ from the stress model for reasons to become apparent shortly. As points of reference, four black dots on each panel indicate the poles for those planes optimally oriented to the stress field if $\mu = 0.6$. The large red dots are poles to the two well-oriented fault planes (Fig. 2a-c) and for the more poorly oriented plane (Fig. 2d-f). Finally, contour lines of constant *SNR* = 0.4 (grey) or 0.8 (purple) are included representing these two frictional limits according to Eqn. 1.

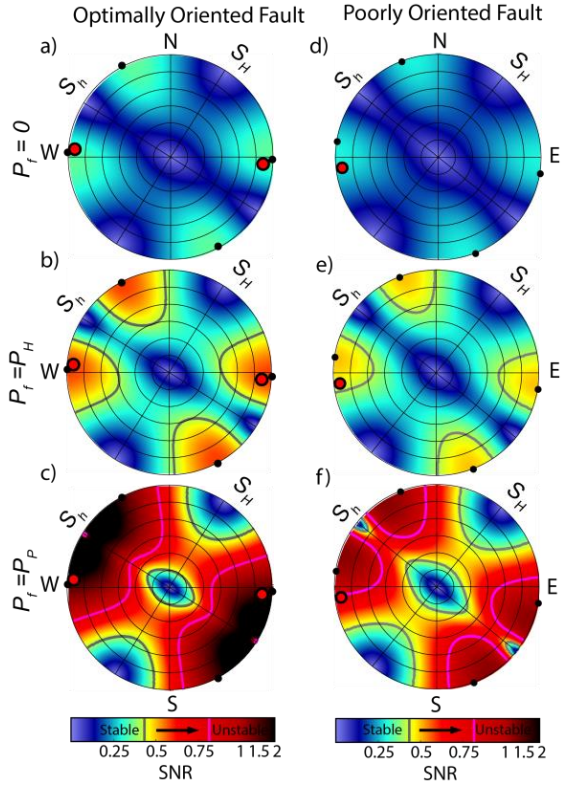

**Figure 2.** Stereonet plot with *SNR* shown in nonlinear colormap for the poles of any arbitrarily oriented faulting planes in the at the epicenters of the well-oriented $M_W$ 3.6 and 4.1 events with the Andersonian model stresses $S_h$ = 65 MPa; $S_V$ = 84 MPa, $S_H$ = 124 MPa with $\phi$ = 41° using a) $P_f = 0$, b) Hydrostatic $P_H$ = 33 MPa, and c) $P_f$ = measured $P_P$ 62 MPa; and for the poorly oriented $M_W$ 3.9 event d) $P_f = 0$, e) Hydrostatic $P_H$ = 33 MPa, and f) $P_f$ = measured $P_P$ 57 MPa. See text for details.

According to Eqn. 1, the fault is unstable if $SNR > \mu$. To interpret, the areas of the stereonets with values of *SNR* greater than the slip criterion value of $\mu$ chosen represents the set of unstable fault planes. Consequently if $P_f = 0$ (Fig. 2a,d) *SNR* reaches a maximum of ~0.3 indicating that the faults could slip only if friction is low; and suggests that the faults would likely be stable. This situation persists if $P_f$ (Fig. 2b,e) is at the more normal hydrostatic gradient with low values of *SNR* ~0.5 expected on the faults. However, the situation changes significantly with larger $P_f \approx P_P$ (Fig. 2c,f) where *SNR* is larger on both fault planes. Indeed, $SNR \geq 2$ for the optimally oriented faults (Fig. 2c); barring unexpectedly high friction or cohesion these faults could not remain stable if the formation pore fluid pressure $P_P$ was active on their planes. Alternatively, for the more poorly oriented $M_w$ 3.9 event (Fig. 2f) $SNR \approx 0.8$.

Within the historical record these faults to the best of our knowledge have appeared to remained clamped (i.e., did not detectably slip). This suggests that the fluid pressure $P_f$ active in the immediate vicinity of the slip planes cannot be the same as the highly overpressured $P_P$ measured from borehole testing within the Duvernay Formation One possible explanation for this is that the planes of weakness are conduits providing transmissive migration pathways for fluids generated within the Duvernay Formation to the overlying and more normally pressured conventional siliclastic reservoirs. That such pathways may exist is not unexpected given that the Duvernay Formation is believed to be the source rock for much of hydrocarbons within the prolific Western Canada Sedimentary Basin [*Stoakes and Creaney*, 1985]. Fluid pressures within such zones could be relieved via along zones of hydraulic connectivity either continuously or possibly via fault valving mechanisms [*Sibson*, 1990].

As might be expected, regardless of $P_f$ the two earthquakes ($M_W$ 3.6 & 4.1) occur on planes whose poles are close to the maximum *SNR* in Fig. 2a-c, , this seeming agreement warrants further examination. On the one hand, the orientations of the earthquake slip planes are independently given by the focal mechanisms as constructed from the events' radiation patterns. On the other, because the stress state is Andersonian with $S_V$ vertical the orientation of the maximum *SNR* orientations in the strike-slip environment is entirely controlled by $\phi$, which is also completely independently obtained from examination of borehole image logs. Closer examination of Fig. 2 shows that here the fault planes strike at angles $\theta = 35°$ and $37°$ from $\phi$ for the $M_W$ 3.6 and 4.1 events, respectively. This is worth pointing out because the azimuth of focal mechanism *p*-axes, which by definition are 45° from the fault plane, are often taken as a proxy for the stress directions. The smaller angle between the fault plane and $\phi$ is consistent with frictional constraints.

## 4. *Triggering of fault slip with increasing $P_f$*

Carrying out more detailed explorations of the influence of $\mu$, $P_f$, and $C$ on a case by case basis using only slip-tendency plots of Fig. 2, or equivalently Mohr stress diagrams, is cumbersome [e.g., *Lele et al.*,

2017]. Instead, in Fig. 3 we plot *SNR* as a function of $P_f$ directly on the fault planes for the optimally oriented $M_W 4.1$ (Fig. 3a) and the poorly oriented $M_W 3.9$ (Fig. 3b) using their respective stress tensors (Table S2) but assuming either $C = 0$ (red band) or $C = 5$ MPa (green band). Within Fig. 3 at any given $P_f$, the vertical thickness of the band accounts for the uncertainties of the stress tensor at that hypocenter (Table S2) and is primarily controlled by the larger distribution of expected $S_H$ magnitudes. The same plots for the remaining events (Figs. S6.1 to S6.9) are provided in the supplementary materials. The vertical dashed lines in Fig. 3 delineate fluid pressures of the hydrostat $P_H$, the measured Duvernay Formation pore pressure $P_P$, and the measured minimum horizontal compressive stress $S_h$. This latter pore pressure represents an upper limit as once it exceeds $S_h$ natural hydraulic fracturing would be expected. Typically, workers assume $\mu$ ranges from 0.6 to 1.0 in the upper crust [*Byerlee*, 1978]. Here we use the range $0.4 \leq \mu \leq 0.8$, constrained by a variety of experimental friction tests (Table S3) on rocks similar to the clay-poor and stiff Duvernay Formation [*Ong et al.*, 2016], although we reiterate that there are no direct measurements of the frictional properties of Duvernay Formation rock currently available to our knowledge. Similarly, no direct measures of fault cohesion (*C*) exist; justification for use of these low values is provided in the supplementary materials.

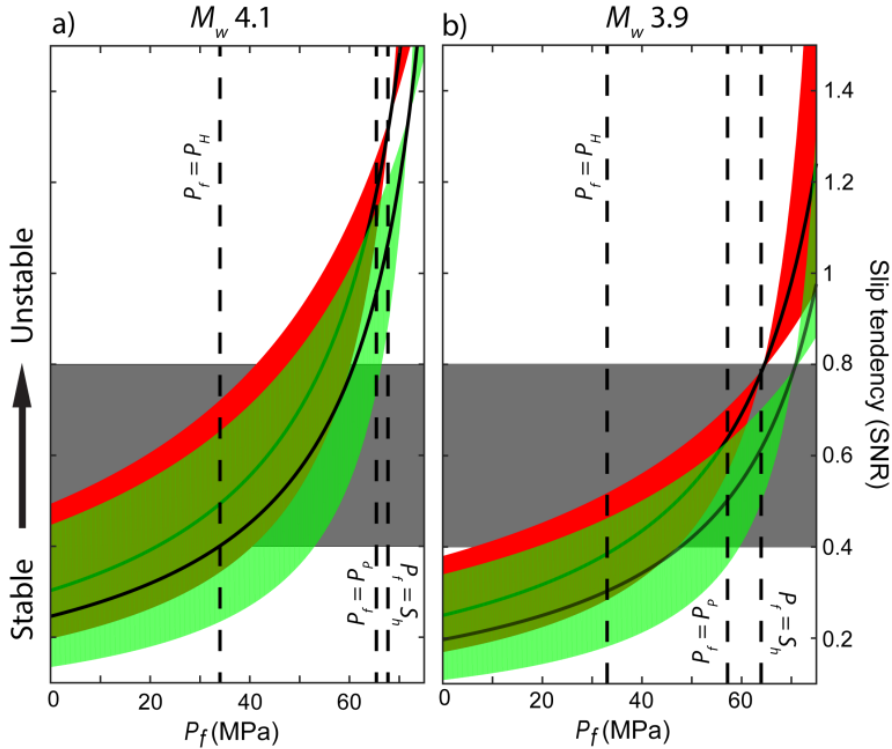

**Figure 3.** Increasing slip tendency (*SNR*) of the fault with rising $P_f$ for a) $M_W$ 4.1 and b) 3.9 events. The *SNR* of the earthquake's faulting plane at different fluid pressures. The red and green regions represent the SNR distributions setting, respectively, for $C = 0$ and $C = 5$ MPa. The width of the stripe represents the uncertainty of *SNR* due to different $S_H$, and the black lines show the calculation with the most confident value for $S_H$. Gray box highlights the range of *SNR* (0.4- 0.8).

Both fault planes are likely stable if $P_f = P_H$ remains at the normal hydrostat. The poorly oriented fault plane (Fig. 3b), too, probably remains stable even at pressures exceeding the ambient $P_p$ (57 MPa). In contrast, at the ambient $P_P$ (62 MPa) the optimally oriented fault plane (Fig. 3a) is already unstable. The historical quiescence of this fault could be interpreted variously to mean that *i)* it is characterized by unexpectedly high values of $\mu$ or of *C*, or *ii)* that the $P_f$ naturally active on the fault is significantly less than $P_p$. The stabilities on all 11 fault planes are further explored in Fig. 3a by calculating their individual *SNR* distributions using a Monte-Carlo procedure (see Supplementary Material) that incorporates the uncertainties associated with the three stress magnitudes and depths, and over expected ranges of varying $\mu$ and *C* for two extreme pore fluid pressure cases with $P_f = 0$ (blue) and $P_f = P_P$ (green). Each distribution is shown as a box and whisker format plotted versus the local angle difference $\psi = S - \phi$ between fault strike *S* and the $S_H$ direction (Table S1). $\psi$ can be considered as a proxy measure of how well a given fault plane is oriented with respect to slip with the vertical gray band indicating the range of optimal $\psi$ orientations of 26° to 34° corresponding to the range of $\mu$ between 0.8 to 0.4. The *SNR* distributions of these earthquakes are mostly below 0.4 in the unrealistic case with $P_f = 0$ (Fig. 4a) reinforcing the expectation that for this extreme case all of the faults would likely remain stable. In contrast, if the faults were perfectly hydraulically connected to the Duvernay Formation such that $P_f = P_P$, almost all the *SNR* distributions shift to values well in excess of 0.8 again indicating that nearly all the faults would be unstable at ambient $P_p$ within the reservoir.

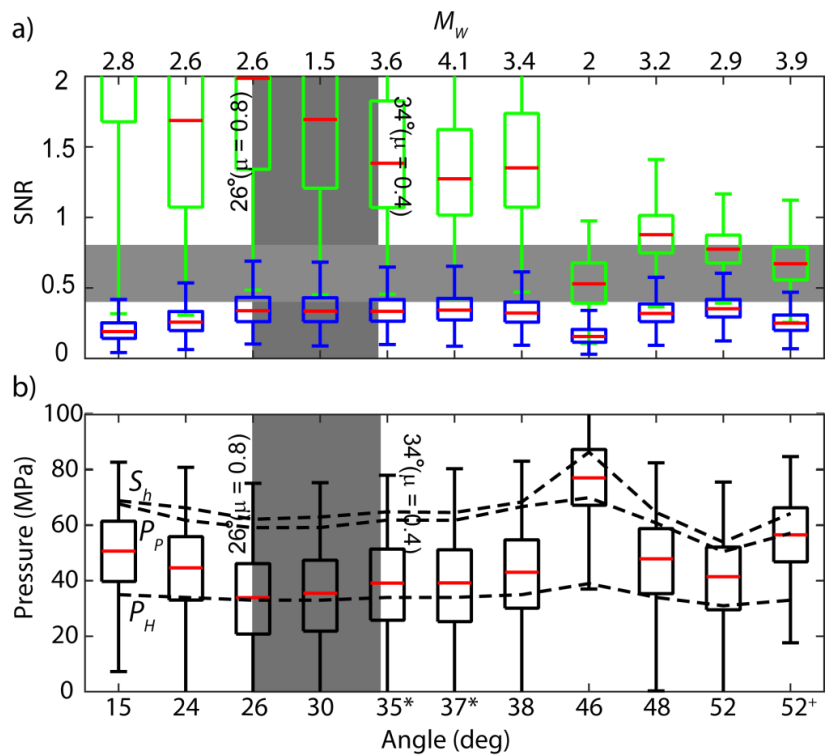

**Figure 4.** Monte-Carlo calculations versus angle $\psi$ of a) *SNR* distributions on the fault planes calculated assuming $P_f = 0$ (blue) and $P_f = P_P$ (green), boxes indicate the 25th and 75th percentile limits of the probability density functions, and the red lines indicate the most probable SNR value for each case, plotted versus the angle $\psi$ bisecting each events local $S_H$ direction $\phi$ and the fault plane strike. b) The distributions fluid pressures needed to activate the faults, boxes indicate the 25th and 75th percentile limits of the probability density functions, and the red lines indicate the most probable critical $P_f$ value. Black dashed lines represent the fracture closure pressure (equal to $S_h$), measured virgin $P_P$ and depth-dependent normal hydrostatic pressure $P_H$. Shaded gray zones in both panels indicate suggested *SNR* and optimal orientation angle ranges.

Given that the faults appear unstable under the expected virgin formation $P_P$, distributions of the greatest allowable values for $P_f$ that are necessary to maintain stability (i.e., $SNR \leq \mu$) were further explored using a Monte-Carlo approach for restricted ranges of $\mu$ and $C$ with the individual stress states. For most of the cases, the most probable $P_f$ (Fig. 4b, see supplementary material for methods) required to initiate slip is slightly greater than the local $P_H$ but significantly less than the local $P_P$. This again indicates that the faults are likely not stable under the high ambient formation $P_P$ and suggests that the virgin $P_f$ acting on the faults must be lower.

During hydraulic fracture stimulations, the fluid pressure required to drive a hydraulic fracture into a rock mass must at least exceed the fracture closure pressure (here $S_h$), a value that in practice is readily exceeded by the actual pressures measured in the wellbore $P_w(t)$ at the injection point during stimulation. Consequently, fluid pressures sufficiently high to trigger slip are present within the system during stimulation, although the actual fluid pressures delivered to the fault through the induced fracture network from the borehole remain highly uncertain. Non-double couple focal mechanism components anecdotally suggest fluid inflow into the fault during the earthquake [*Wang et al.*, 2018]. Reportedly, fractures stimulated during the HF well completion extend laterally no more than 400 m with excursions to nearly 900 m from the injection point [*Wilson et al.*, 2018]; and this may provide some insight on the distance fluid pressures could be transmitted during an HF stimulation. That said, model-derived estimates depend on having knowledge of many largely unconstrained factors controlling fracture and fault geometries and fluid transport. Poroelastic stresses, too, may contribute to the stress state on the fault planes at initiation uncertain in these formations but numerical [*Chang and Segall*, 2016] and analytical [*Segall and Lu*, 2015] simulations suggest their influence on the *SNR* is a small fraction relative to $P_f$.

In Fig. 4, the local angle difference $\psi$ is intentionally chosen as the independent variable to emphasize that the planes of weakness are not necessarily all optimally aligned with the stress field. The suggested range $0.4 \leq \mu \leq 0.8$ delimits the optimal range $34° \geq \psi \geq 26°$ highlighted in Fig. 4; which almost all of the events fall outside of. This observation could have multiple interpretations. The most likely explanation is that more optimally oriented planes of weakness are absent at these locations; slip instead occurs on those pre-existing faults that are closest to unstable but not perfectly aligned with the stress field. Another interpretation is that the slip does occur on the real fault planes that are optimally oriented, but the FM planes or values of stress (retaining the Andersonian assumption) are uncertain.

5. **Conclusion**

The slip-tendencies along faults activated by hydraulic fracture stimulations in a localized area of NW Alberta, Canada were analyzed using a recently developed quantitative model for the full stress tensor and the formation pore fluid pressures $P_p$ within the highly over-pressured Duvernay Formation. Assuming reasonable ranges for fault friction and cohesion, nearly all of the slip-planes studied would be unstable at the measured ambient formation pore fluid pressures $P_p$. This instability persists although most of the slip-planes are not expected to be optimally oriented with respect to the prevailing stress directions. That this area was historically aseismic prior to hydraulic fracturing operations, however, indicates that the natural fluid pressures within the fault zone must be lower unless unexpectedly large frictions or cohesions exist; Monte-Carlo simulations suggest that generally, the most probable critical fluid pressures lie closer to the normal hydrostatic pressure. As hydraulic fracturing stimulations generally attempt to maintain fluid pressures above $S_h > P_p$, the potential to convey in excess of a critical pressure to the surrounding formation exist, although actually quantitatively estimating the critical pressure is difficult. One additional comment arising from is that including even a modest cohesion $C$ does affect the *SNR* value noticeably; and while omitting $C$ may be useful within the context of engineering risk assessment better understanding this phenomena warrants further study. The inferred lower pressures within the faults suggest that they may serve to provide conduits for migration of hydrocarbons out of the low permeability Duvernay Formation to the overlying siliclastic formations and may be consistent with the critically stressed crust hypothesis [*Townend and Zoback*, 2000]. The results here highlight the challenges confronting researchers hoping to understand the physics of earthquake rupture by artificially initiating fault slip [*Savage et al.*, 2017].

## 6. Acknowledgement

The authors thank the Alberta Energy Regulator for allowing the development and publication of this work. Earlier components of the work were supported by Helmholtz-Alberta Initiative and NSERC Discovery Grant. Supporting figures, tables and data are available in the supplementary information.